\documentclass[12pt]{article}
\input epsf.tex

\usepackage{latexsym}
\usepackage{graphicx}
\usepackage{amsmath,amstext}
\usepackage{comment}
\numberwithin{equation}{section}
\newcommand{\be}{\begin{equation}}
\newcommand{\ee}{\end{equation}}
\newcommand{\benn}{\begin{equation*}}
\newcommand{\eenn}{\end{equation*}}
\newcommand{\bea}{\begin{eqnarray}}
\newcommand{\eea}{\end{eqnarray}}
\newcommand{\bean}{\begin{eqnarray*}}
\newcommand{\eean}{\end{eqnarray*}}

\begin{document}
\begin{titlepage}
\begin{center}

\vskip0.2in

{\Large \bf Minimal Deflected Anomaly Mediation}

\vskip 0.3 in

Linda M. Carpenter$^1$

\vskip0.2in
 
\emph{$^1$ Johns Hopkins University\\Baltimore MD}

\begin{abstract}
I provide a minimal method for generating a mu term and positive
slepton masses within the framework of
anomaly mediation. I introduce a new operator involving a hidden
sector U(1) gauge field which is then canceled against a Giudice-Masiero-like $\mu$ term.  No 
new flavor violating operators are allowed.  This 
procedure produces viable electroweak symmetry breaking
 in the Higgs sector. 
Only a single pair of new vector-like messenger fields is needed to correct the slepton masses by deflecting them from their anomaly mediated 
trajectories.  This model is minimal but it requires fine-tuning.

\end{abstract}
\end{center}
\end{titlepage}

\section{Introduction}
Anomaly mediated supersymmetry breaking provides a flavor blind, UV insensitive, predictive method of supersymmetry breaking.  In this way it almost
 fulfills the long wish
list of things we would want from the perfect model of supersymmetry breaking \cite{3}\cite{13}.  However, it has two main problems; the first is that it generates
 negative slepton mass 
squareds, the second is the $\mu$ problem, or the generation of a weak scale mass for the Higgsinos. Within the framework of anomaly mediation, we can generate a Giudice-Masiero-like  $\mu$ term that doesn't require a singlet.  However,
 the resulting $B$ term will be a loop factor too large to facilitate proper electroweak symmetry breaking.  This leads one to ignore this mu term and search for a fix elsewhere. 

Clearly we need extra sources of SUSY breaking to create a viable model.  One can generate soft masses for fields through generic hidden sector supersymmetry breaking; for example, contact terms with hidden sector fields that have $F$ term vevs
\bea
\int d^4 \theta \frac{XX^{\dagger}}{M^2}QQ^{\dagger}.
\eea
However, these operators may have unsuppressed flavor violation.  Instead we may add a new source of supersymmetry breaking, a U(1) gauge field which 
acquires a $D$ term vev in the hidden sector.  This doesn't allow for direct contact terms with the scalar fields $Q$ which would contribute to flavor 
violating processes at leading order.  This was explored in ref \cite{17}.
   In the MSSM, a single new operator can be generated which is an additional $B\mu$ term for the Higgs fields, $\int d^2\theta W^{'}W^{'}H_u H_d$ \cite{10}.  With this new term, we need not throw out the Giudice-Masiero-like $\mu$ term but instead can keep it and tune the $B\mu$ term against it to get the correct electroweak vev.  This mechanism is a module which can be used in conjunction with several different methods of fixing anomaly mediation's slepton problem, for example the addition of Fayet-Iliopoulos D-terms as in refs \cite{4}\cite{9}.

Another such method is the addition of extra chiral superfields, or messengers, in a vector-like representation.  Such fields have a nonsupersymmetric mass threshold 
with mu-like and $B$-like terms generated by a Kahler potential operator, exactly like the Higgs fields.  This mass threshold pushes the gauginos off of 
their anomaly mediated trajectory, and adds to the
 scalar masses at two loops.  Multiple sets of these vector-like fields are typically required to counteract the negative slepton masses from anomaly 
 mediation, but with the enhanced $B$ term the deflection becomes much greater.  Thus
    the slepton masses are driven positive with 
  only one set of extra chiral superfields.  If the single messenger is in a complete SU(5) representation, then perturbative unification is easily preserved. Since the mass thresholds are set at only $10$ TeV, UV insensitivity is preserved down to that scale.
   With only a small number of new parameters, the theory remains predictive and viable, but the trade off is for moderately heavy fine tuning.   
   
This paper has the following outline: Section 2 provides an overview of anomaly mediation and discusses how the new B term arises from the addition of the new U(1).  Section 3 
discusses electroweak symmetry breaking and minimal fixes to the slepton problem.  In Section 4 viable  spectra are produced and discussed, and Section 5 is
conclusions.

\section{SUSY Breaking}    

To allow anomaly mediation to dominate and prevent arbitrary flavor violation, one must forbid contact terms between the MSSM and the hidden sector.  One way this can be achieved is by
using a 5-D setup
with two boundaries separated by the extra dimension.  The standard MSSM fields inhabit one boundary and hidden sector fields inhabit the other, with only gravity propagating in the extra dimension.  Supersymmetry is broken on the hidden sector boundary and is communicated to the visible sector as the F term of the conformal compensator
  $\Phi= 1+ {\theta}^2 m_{3/2} $.  The conformal compensator is the spurion of broken scale invariance and appears in the Lagrangian next to any explicit mass scale.  Thus after rescaling fields and regulating, the conformal compensator appears with $\Lambda$, the cutoff of the theory. The Lagrangian for the set of fields $Q_i$ is 
\bea
L = \int d^4 \theta[Z_i(\frac{\mu}{\sqrt{\Phi\Phi^{\dagger}}\Lambda})Q_i^{\dagger} e^{-2V} Q_i]+\int d^2\theta \frac{1}{2}g^{-2}(\frac{\mu}{\Phi\Lambda}) Tr[W^{\alpha}W_{\alpha}]+h.c. \nonumber
\\ - \int d^2 \theta [ \lambda_{ijk} Q_i Q_j Q_k +\Phi m_{ij} Q_i Q_j + {\Phi}^2{v_i}^2Q_i] +h.c. 
\eea
where $\mu$ is the renormalization scale.
Expanding the functions of $\Phi$  in terms of ${\theta}^2$  yields gaugino masses

\be\label{eq:gauginomass}
m_{\lambda_i}=\frac{\beta(g_i)}{g_i}m_{3/2}.
\ee
and scalar masses and trilinear terms

\bea\label{eqn:ammasses}
m^2_i = -\frac{1}{4} \dot{\gamma}m_{3/2}^2 & & A_{ijk}= -\frac{1}{2}(\gamma_i
+\gamma_j +\gamma_k) m_{3/2}
\eea

For mass thresholds that are supersymmetric,  the soft masses at low energy depend only on the value of the $\beta$ functions at that energy. Thus we call the theory UV insensitive.  Also, a single scale
 parameter $m_{3/2}$ sets the masses of the superpartners, therefore this form of SUSY breaking is highly predictive.  Only scalars that participate in the strong force 
 get large positive contribution to their anomaly mediated masses.  The sleptons 
 have only non-asymptotically free couplings, thus due to their $\beta$ functions their anomaly mediated mass squareds are negative.    Any 
  complete model employing anomaly mediation will have to address this issue.  

On a more positive note, we can generate a mu term using a Giudice-Masiero-like operator without introducing a singlet \cite{2}\cite{12}.  We may write down a term in the Kahler potential
\bea
\lambda_h \int d^4 \theta \phi^{\dagger} \phi H_{u} H_{d}
\eea
We can then rescale the fields $H \rightarrow \phi H$ to get
\bea
\lambda_h \int d^4 \theta \frac{\phi^{\dagger}}{\phi} H_{u} H_{d}.
\eea
Inserting for the conformal compensator $\phi = 1 +m_{3/2} \theta^2$ and expanding
gives 
\bea
\lambda_h \int d^2 \theta m_{3/2} H_{u} H_{d} - \lambda_h m_{3/2}^2 h_u 
h_d
\eea
The first term is a mu term for the Higgsinos.  The second term is a B term for the scalar Higgs. The scale $m_{3/2}$  is $\sim 10$ TeV, so that with a choice of small
 coupling, $\lambda_h \sim \frac{1}{16\pi^2}$, we may have a mu term that is of order the weak scale.  However in this case, the B term is of order $\frac{m_{3/2}^2}{4 \pi}$.  With such a $B$ term, the 
 dynamics of electroweak symmetry breaking would generate a Higgs vev at the scale $m_{3/2}$ which is a loop factor too large.  A new mechanism is needed to create a
 viable $\mu$ term.

In this model, there is one additional source of SUSY breaking.  We have a U(1) gauge symmetry; in our 5-D setup it propagates in the bulk and is broken on the
hidden sector boundary.  The gauge field gets a $D$ 
term vev by some dynamical mechanism in the hidden sector.  Since we want a $D$ term that is the same size as the overall SUSY breaking scale, we may deduce that
the $D$ term vev is itself closely connected to, and possible even required for supersymmetry breaking.  For examples of such models see refs. \cite{21},\cite{10}.  With the addition of this extra U(1) gauge field, there are only two new operators that we may write down
 with all Lorentz and gauge indices contracted. 

One operator is
\be
\frac{c_h}{{M}^2} \int d^2 \theta W^{'} W^{'}  H_u H_d .
\ee
When the $D$ term is set to it's vev this term becomes
\be
 {c_h}{m_D^2}H_u H_d
\ee
with $\frac{D}{M_{cut}} \equiv m_{D}$, and $m_{D} \sim O(m_{3/2}) $.
  This is an additional $B$ term which we can cancel against the Giudice-Masiero-like  $B$ term.  The entire $B\mu$ term is now an adjustable parameter which need
  not be of order $\lambda_h m_{3/2}^2$.  A one percent tuning between the scales $\lambda m_{3/2}^2$ and $c_h m_D^2$ is enough to give the correct electroweak
  physics.  The new 
  operator only adds one more parameter to the theory, the scale $c_h m_{D}$, so the mu problem can be solved while maintaining an economy of 
  parameters.  

We generate one additional operator that could potentially cause problems in this model:

\be
\int d^2 \theta W_Y W^{'}
\ee
which is a tadpole term for the hypercharge $D$ term.  One cannot forbid this new term through boundary conditions as the new U(1) lives in the bulk and must communicate with at least some 
fields on the standard model brane.  However, if we choose not to write down this operator in the first place it will not be generated by radiative corrections.  For those who
 worry that whatever operator can be written down must be included, one could try to forbid this term with symmetries. For example one can introduce charge conjugation under which only $W^{'}$ is odd, and which is
   broken only on the hidden sector boundary.  Once the extra dimension is integrated out, operators with charge conjugation violation will be communicated to the standard 
  model brane but will be suppressed at the least by powers of the SUSY breaking scale over the compactification radius.        

Scalar masses for squarks and sleptons cannot be generated through direct contact terms with the hidden sector gauge field.  Holomorphy 
 prevents us from writing such a term in the superpotential.  Instead the
  lowest dimension mass term we may write is $\frac{1}{M^6}\int d^4 \theta W^{'}W^{'}W^{'\dagger}W^{'\dagger}QQ^{\dagger}$, which is highly suppressed and not generated by any divergent diagrams.

There is a correction to the Wino mass and slepton mass due to loop effects from the Higgsinos.  The Wino mass is corrected at one loop in a way similar to gauge mediation (\cite{14}\cite{15}\cite{21}), with the overall contribution going like 
\be
\frac{{g_y}^2}{16 {\pi}^2} \frac {B \mu}{\mu}
\ee
Generically this correction is of order $\frac{1}{16 {\pi}^2} \mu$.  The masses of the sleptons are corrected at two loops with
diagrams involving the Winos, Higgs' and Higgsinos.  These corrections are  of order 
\be
m_{sl}^2 \sim {\frac {B \mu}{\mu}}^2 \left( \frac{{g}^2}{16 {\pi}^2}\right)^2
\ee
For a weak scale $\mu$ term however, this correction is not nearly enough to fix the negative slepton masses from anomaly mediation; additional structure will
 have to be added to complete the model.

\section{Deflected Anomaly Mediation}

One method of fixing the slepton problem involves changing the anomaly 
mediated trajectories of the scalars and gauginos by adding  chiral superfields in a vector like representation \cite{6} \cite{2} \cite{1}.   These fields are known
as messengers.   The messengers $\Psi$ will have a term in the Kahler potential $\int d^4 
\theta \phi^{\dagger} \phi \Psi \overline{\Psi}$. This can be rescaled as $\int d^2 
\theta \frac{\lambda_1 m_{3/2}}{\phi} \Psi \overline{\Psi}$ giving rise to mu-like and $B$-like 
terms analogous to those of the Higgs sector.  The effect of the new fields is to change the anomaly mediated 
contribution of the gauginos in a way similar 
to gauge mediation.  Like the example of the Higgs loops contributing to 
gaugino masses in Section 2, the new messenger loops contribute a mass 
roughly the size $\frac{g^2}{16\pi^2}\mu$. However, in this 
case there are no further constraints from electroweak symmetry breaking 
requiring the $\mu$-like term, which we will call $M$, to be of order the weak scale.  M can then be it's more natural value
of order $m_{3/2}$.  The gaugino contribution is now free to 
be of order $ \frac{1}{16\pi^2}m_{3/2}$.  The slepton masses are then corrected as the running is pushed off of its anomaly mediated trajectory.  

We will first derive the masses of the gauginos and scalar superpartners which are deflected by the $\mu$-like and $B$-like thresholds of the Kahler potential operator alone.  This formalism is worked out in refs \cite{1}\cite{6}\cite{16}\cite{18}.  We define the mu-like term $M \equiv \lambda m_{3/2}$ and the $B$-like term $F \equiv -\lambda m_{3/2}^2$. Thus the mass of the messengers is defined to be  $X \equiv M +\theta^2 F$.  
We can look directly at the gauge couplings and extract the mass of the gauginos.  Note that we will be looking at non-holomorphic gauge couplings.  The holomorphic gauge couplings may be expressed as functions of the real gauge couplings as discussed in ref \cite{1}.  For the mass threshold $X$ we see 

\bea
\alpha^{-1}(\mu, X)=\alpha^{-1}(\Lambda)+\frac{b-N}{4\pi}ln \left( \frac{XX^{\dagger}}{\Lambda \Lambda^{\dagger}
\Phi \Phi^{\dagger}}\right)+\frac{b}{4\pi}ln \left( \frac{\mu^2}{XX^{\dagger}}\right)
\eea
where $N$ is the number of sets of messengers \cite{1}\cite{7}. The low energy beta function coefficient $b$ is just the beta function coefficient above the new mass threshold
 minus the number of extra chiral superfields, or
\be
b_{UV} = b - N.
\ee

A supersymmetric mass threshold $M$ comes with one power of the 
conformal compensator, as the Lagrangian of Equation 2.1 demonstrates,  which cancels the powers of the $\Phi$ that appear with the cutoff.  In this way, the first term in the equation above contributes nothing to
the gauge coupling. The only contribution comes from the second term, which when we expand gives us just the contribution we expect from anomaly mediation.  Thus the high energy running is wiped out and only the low energy anomaly mediated
trajectory remains. In this way the low energy theory is insensitive to the UV physics.  However, in our case we have a nonsupersymmetric mass scale $X = \lambda\frac{m_{3/2}}{\Phi}$ so the gauge coupling becomes  
\bea
\alpha^{-1}(\mu, X)=\alpha^{-1}(\Lambda )+\frac{b-N}{4\pi}ln\left( \frac{\lambda^2 
m_{3/2}^2}{\Lambda ^2
{(\Phi \Phi^{\dagger}})^2}\right) +\frac{b}{4\pi}ln\left( \frac{\mu^2\Phi \Phi^{\dagger}}{\lambda^2 m_{3/2}^2}\right),
\eea
If we expand in terms of $\theta^2$ contained in the conformal compensator, $\Phi \rightarrow 1+\theta^2 m_{3/2}$, we see that the gaugino masses go like
\be
m_{\lambda_i}= \frac{{\alpha_i}^2}{4 \pi} (b_i-2N)m_{3/2}.
\ee
Since the gaugino masses have changed, the running of the scalar masses is deflected from its anomaly mediated trajectory.

For the scalars, we may expand the wave function renormalization in terms of $\theta^2$. After rescaling fields
we can extract the $\theta^2 \overline{\theta^2}$ component, just like we do in anomaly mediation.
This gives scalar masses 
\be
m_s^2=-\frac{1}{4}m_{3/2}^2\left( \frac{\partial^2}{\partial^2 ln\mu} - 
d\frac{\partial^2}{\partial^2 ln X}\right)^2 ln(Z)
\ee
with
\be
Z_i(\mu, X) = Z_i(\Lambda)\left( \frac{\alpha(\Lambda)}{\alpha(X)} \right)^{\frac{2c_i}{b_i-N}}\left( \frac{\alpha(X)}{\alpha(\mu)} \right)^{\frac{2c_i}{b_i}}.
\ee

We define the deflection parameter $d$ such that 
$\frac{F}{M}-m_{3/2}=dm_{3/2}$.  The square of the first term is just the anomaly mediated contribution, 
and the running from the threshold is contained in the rest.  For the case 
of a supersymmetric mass threshold we have
$\frac{F}{M}=m_{3/2}$ or $d=0$.  This just reproduces the anomaly 
mediated mass for scalars.  We see that $d$ parameterizes the difference 
between the anomaly mediated mass threshold and the new mass threshold we 
have added.  Differentiating, we get a scalar mass of

\be
m_s^2 = \Sigma_i \frac{2c_i b_i}{{16 \pi}^2}\left( \left( \frac{N}{b_i}\alpha_{\Psi}^2 - 
\frac{N^2}{b_i^2}[\alpha_{\Psi}^2-\alpha_{\mu}^2]\right)d^2 +
2\frac{N}{b_i}\alpha_{\mu}^2 d \right)m_{3/2}^2+\Sigma_i \frac{2c_i b_i}{{16 \pi}^2}\alpha_{\mu}^2 m_{3/2}^2
\ee
In our case, $\frac{F}{M}=-m_{3/2}$ so we have $d=-2$.   
This gives positive slepton masses for $N \ge 5$.

However, in this case there is one additional operator that can be written down, a new $B$ term that comes from 
the operator $\int d^2 \theta c W^{'} W^{'} \Psi \overline{\Psi}$, in exact analogy to the Higgs sector.  This is an addition 
of $c m_D^2$ to the scalar messenger mass squared. We will now rederive the gaugino and scalar masses with this new contribution.
 We define $c m_D^2 \equiv -\kappa m_{3/2}^2 \equiv B$.  Now we have $X \equiv M + \theta^2 F + \theta^2 B$.  
so $\alpha^{-1}$ becomes

\bea
\alpha^{-1}(\mu, X)&=&\alpha^{-1}(\Lambda)+\frac{b-N}{4\pi}ln\left( \frac{(M+\theta^2 F+\theta^2 B)(M+\overline{\theta^2}F+\overline{\theta^2}B)}{\Lambda^2\Phi \Phi^{\dagger}}\right) \nonumber \\
&+&\frac{b}{4\pi}ln\left( \frac{\mu^2}{(M+\theta^2 F+\theta^2 B)(M+\overline{\theta^2}F+\overline{\theta^2}B)}\right)
\eea
and the gaugino masses become
\bea
m_{\lambda_i}= \frac{\alpha_i}{4 \pi} (b_i-2N) m_{3/2}-\frac{\alpha_i}{4 \pi}N\frac{\kappa}{\lambda}m_{3/2}. 
\eea

The the scalar masses are still given by Equations (3.5) and (3.7)
but we have now changed the deflection parameter to $d = -2 - 
\frac{\kappa}{\lambda}$ Notice that $dm_{3/2}=\frac{2F+B}{M}$. We may now use $\kappa$ to adjust the $B$ term such that the number of necessary copies of messenger multiplets is only one.  
To do this we need the coupling $\kappa$ to be greater than 1.   In this case we 
may think of the greater than 1 coupling arising from picking the cutoff to be $\frac{M_{cut}}{\sqrt{\kappa}}$.  Thus instead of having 
a cutoff at, say,  the Planck scale, the theory requires
new physics at $\frac{M_{Pl}}{a few}$.  Notice that at the threshold in the limit that $B$ gets very large and the deflection parameter becomes very negative, and the dominant contribution to scalar masses comes from our new source of SUSY breaking.

\section{Electroweak Symmetry Breaking and Spectrum}

The Higgs potential for the neutral scalars looks exactly like that of the MSSM, 

\bea
V &=& (\mu^2 + m_{Hu}^2)|H_{u}|)^2 + ( \mu^2 + m_{Hd})|H_d|)^2 -((\frac{\mu^2}{\lambda_h}- c m_{D}^2)H_u H_d + h.c.) \nonumber \\
&+& \frac{1}{8} (g_2^2 +g_y^2)(|H_u|^2-|H_d|^2 )^2  
\eea
The conditions for finding the minimum are 
\be
{m_Z}^2 = -\frac{m_{Hu}^2-m_{Hd}^2}{\cos{2\beta}}-(m_{Hu}^2 + m_{Hd}^2 + 2\mu^2)
\ee
 
 \be
 \sin{2\beta} = -\frac{2B}{m_{Hu}^2 + m_{Hd}^2 + 2\mu^2}.
 \ee

  It was previously noted that for $\mu = \lambda m_{3/2} \sim$ weak scale with $B$ fixed at $\lambda m_{3/2}^2$, no solution exists, as we can see from the second condition \cite{8}\cite{5}.   However, in this new model with an adjustable $B$ term there are two regions of parameter space where we find viable solutions.  One  region has large values of the mu term, $\mu \sim m_{3/2}$ with a coupling $\lambda_h$ greater than 1. Such values could possibly fix the slepton mass problem without adding any messengers at all.  In this case, the correction to the
slepton mass squareds would be of order $\left( \frac{1}{16\pi^2}\right)^2 \frac{B^2}{\mu}$. From the first condition we see that these solutions are only valid for 
$\cos{2\beta} \rightarrow 0$, or tan$\beta \rightarrow 1$.  Looking at the second equality we see that we must then have $\mu^2 \sim B$.  Such points give positive slepton masses for the
choice $\lambda_h$ of order 5.  All scalar masses get large contributions from the deflection, therefore we may shift down the entire spectrum by picking a small
$m_{3/2}$, just at 10 TeV or so.  Unfortunately, such points are significantly fine-tuned. In addition, such a low tan$\beta$ requires a large top Yukawa coupling
that gets a Landau pole far below the unification scale.  These points allow a very minimal model with a single new parameter, but the fine-tuning and top Yukawa
problems leads us to search for solutions in other regions of parameter space. 

Instead we may pick parameters such that the $\mu$ term is of order the weak scale. This will allow large values of tan$\beta$.  In order to satisfy equation (4.2)  we must pick $\mu^2$ such that it cancels the large anomaly mediated Higgs masses down to the Z mass. This requires a tuning of a little better than one percent.  We may then look at equation(4.3) and see that  we must cancel the two large contributions to the $B$ term down to the scale $\mu^2$.  This is a tuning of a little worse than one percent.  So we see that 
parameters in the Higgs sector involve a total sensitivity of $10^{-4}$, which means that this model is highly constrained but not impossible.  A sample point for this parameter space is given in Table 1.

\begin{table}
\label{tab:points}
\begin{center}
\begin{tabular}{c|c|c|c}
& &  Point 1 & Point 2\\
\hline
inputs:  
&$\frac{m_{3/2}}{16\pi^2}$ & 200 & 100 \\
&$m_D$ & .5$m_{3/2}$ & .5$m_{3/2}$\\ 
Higgs sector couplings:
&$\lambda_h$ & .015 & .026\\
&$c_h$ & .0598 & .1029 \\
Messenger sector couplings:  
&$\lambda$ & $4$ & $20$  \\ 
&$\kappa$& 11.9 & 379 \\
\hline
Higgs Sector:
&$\mu$& 474 &  415\\
&$\tan{\beta}$ & 7.24 & 1.20\\
\hline
sleptons:
&$m_{\tilde{e}_L}$&  446 & 1162 \\
&$m_{\tilde{e}_R}$& 126  & 426 \\
\hline
Gauginos:
&$m_{\tilde{W}}$ & 510 & 936\\
&$m_{\tilde{B}}$ & 407 & 407\\
\hline
Squarks:
&$m_{\tilde{sq}_R}$ & 1235 & 3707 \\
\hline
\hline
\hline
\end{tabular}
\caption{Two possible spectra, all masses given in GeV.}
\end{center}
\end{table}

We must also avoid getting a vev for the scalar messengers $\Psi$.  Thus the mass matrix 
\be
M_{\Psi}^2= m_{3/2}^2\left( \begin{array}{cc}\lambda^2&\lambda+\kappa\\ \lambda +\kappa& \lambda^2 \end{array} \right)
\ee
must 
have positive determinant.  We then have the constraint $\lambda^2 > \lambda + \kappa$, which gives the upper bound on the B term for a fixed $\lambda$. 
We may now calculate the lower bound on $B$; for example with $\lambda \sim 4$ and $\kappa$ = $0$, one extra vector-like multiplet falls short of making the
 slepton 
mass positive by about a factor of 8.  Thus we can estimate $B$, using equation (3.7)
which tells us we must have $B > 1.75 F$.  Unlike those of the Higgs sector, these constraints are not very strict, so the spectrum is viable for a large 
range of couplings.  One feature of the spectrum is that for moderate values of $\lambda$ (less than 10) the Wino mass falls naturally at a few hundred GeV.  The slepton
masses are small, so that the wino is now in the middle of the anomaly mediated mass spectrum.  Thus the right handed sleptons become the LSPs.  Table 1 contains such a point.

 This model differs from minimal anomaly mediation most in its heavier wino mass.  In fact,
 for the moderate values of couplings that yield positive slepton masses, we always have the wino heavier than the bino.  The
 squarks are also much heavier than the minimal model, with masses of a few TeV, due to deflection from their large SU(3) coupling.  We have a heavy and light scalar messenger. The light mass eigenvalue is around 10 TeV.  For these moderate values of the couplings $\lambda$ and $\kappa$ the spectrum looks like some combination of anomaly mediation and low energy gauge mediation, with each contributing effects of the same order; other models employ similar combinations of SUSY breaking sources giving different spectra \cite{19} \cite{20} \cite{11}. 
 
 For the case of very large couplings $\kappa$ and $\lambda$ we
 see that the new $B$ terms dominate the anomaly mediated contributions.  The spectrum looks like gauge mediation, and in fact for
 $N=1$ we see that the bino can be made lighter than the sleptons.  This requires the choice of large coupling, $\lambda > 19$ and $\kappa > 350$.  If we then pick
 a moderate sized $\mu$ term for the Higgs, the bino is the
 lightest superpartner.  At these large values of the messenger couplings the entire spectrum becomes heavy. We can
 drop the entire spectrum by making $m_{3/2}$ smaller.  However, in order to maintain the bino as the LSP, we must pick the couplings in the Higgs sector such that the $\mu$ term is larger than
 the bino mass as we drop $m_{3/2}$.  These larger couplings drive tan$\beta$ lower as was discussed above.  For points with a light overall scale we have tan$\beta \sim 1.2$ which is still not as large as we
 would like.  If we instead pick large values of $m_{3/2} \ge$ 75 TeV, we need not have large Higgs sector couplings to maintain the bino LSP.  We then have
 more viable values of tan$\beta$.  However, the entire spectrum has masses at the TeV scale. The prospect of a bino LSP is exciting, but we would have to accept a scale of new
 physics at $M_{cut} \sim \frac{M_{Pl}}{19}$ and a choice between low tan$\beta$ or a heavy spectrum.

\section{Conclusions}
Anomaly mediated SUSY breaking provides a UV insensitive flavor blind method of generating 
superpartner masses.  A mass term for Higgsinos can be generated employing a Giudice-Masiero-like mechanism,
 however it also generates a $B\mu$ term which is two orders of magnitude too 
large.  A new fix to the $\mu$ problem employs anomaly mediation with an new broken U(1) which generates 
a single new operator, an additional $B$ term.  The new term may be tuned against the
 anomaly mediated  term and generate viable electroweak symmetry breaking.  This mechanism
 adds a minimal number of new parameters to the theory and thus maintains a high level of 
predictivity.  Several fixes to the slepton problem may be used in conjunction with our new $\mu$ term.

One viable model is the addition of N copies of messenger fields which change the 
running of the gauginos and hence the scalar masses at two loops.  By adding an extra B term
for these fields, in analogy to what was done with the Higgs sector, the slepton masses may be 
driven positive with just one extra set of vector-like fields instead of five.   The result $N>5$ destroys the possibility of perturbative unification.  Unification may occur for N=5, however our result of N = 1 allows for unification to be easily preserved. 

The spectrum allows for two interesting but possibly dangerous results.  By sacrificing large values of tan$\beta$ in the Higgs sector, we may achieve positive slepton
masses and viable $\mu$ terms without adding any messenger fields at all.  This is the most minimal model, but it results in perturbative breakdown well below the
unification scale.  Alternatively, keeping the messengers and allowing the cuttoff of the theory to be lowered by a factor or 19 or so from the Plank scale, we produce a spectrum
with the bino as the LSP.  Though this has interesting cosmological implications, we would still need to contend with a heavy spectrum or questionable values tan$\beta$.  The safest points in parameter space produce a spectrum with light sleptons, middle weight winos and possibly a scalar messenger lurking just above the squark masses at 10TeV.  The model generates viable electroweak symmetry breaking, a weak scale mu term, positive slepton mass squareds and a 
viable spectrum while maintaining UV insensitivity, flavor blindness, and a minimal number of 
extra parameters. This is a predictive model, though the parameter space available is very small and thus the theory is fine tuned. 

\vspace{.5in}

{\bf Acknowledgements}
 
I'd like to thank David E. Kaplan for all of his help, and Markus Luty for useful discusions. 

\end{document}